\documentclass[twocolumn,secnumarabic,amssymb, aps, prl,amsmath]{revtex4}

\usepackage{graphicx}
\usepackage{natbib}
\usepackage{amsmath,amssymb,bm}
\usepackage{color}

\newcommand{\Pe}{\mbox{Pe}}
\newcommand{\bs}{\bm}

\newcommand{\dircos}{\chi}
\newcommand{\dirsin}{\sigma}
\begin{document}

\title{Alignment and scattering of colliding active droplets}%

\author{Kevin Lippera}
\affiliation{LadHyX -- D\'epartement de M\'ecanique, CNRS -- Ecole Polytechnique, Institut Polytechnique de Paris, 91128 Palaiseau, France}
\author{Michael Benzaquen}
\affiliation{LadHyX -- D\'epartement de M\'ecanique, CNRS -- Ecole Polytechnique, Institut Polytechnique de Paris, 91128 Palaiseau, France}
\author{S\'ebastien Michelin}
\email{sebastien.michelin@ladhyx.polytechnique.fr}
\affiliation{LadHyX -- D\'epartement de M\'ecanique, CNRS -- Ecole Polytechnique, Institut Polytechnique de Paris, 91128 Palaiseau, France}
\date{\today}

\begin{abstract}

Active droplets emit a chemical solute at their surface that modifies their local interfacial tension. They exploit the nonlinear coupling of the convective transport of solute to the resulting Marangoni flows to self-propel. Such swimming droplets are by nature anti-chemotactic and are repelled by their own chemical wake or their neighbours. The rebound dynamics resulting from pairwise droplet interactions was recently analysed in detail for purely head-on collisions using a specific bispherical approach. Here, we extend this analysis and propose a reduced  model of a generic collision to characterise the alignment and scattering properties of oblique droplet collisions and their potential impact on collective droplet dynamics. A systematic alignment of the droplets' trajectories is observed for symmetric collisions, when the droplets interact directly, and arises from the finite-time rearrangement of the droplets' {chemical wake} during the collision. For more generic collisions, complex and diverse dynamical regimes are observed, whether the droplets interact directly or through their chemical wake, resulting in a significant scattering. 
\end{abstract}

\maketitle

\section{Introduction}
Synthetic microswimmers have recently been a central focus for various scientific communities, whether to mimic life at small scales~\citep{dai2016programmable} or the more general behaviour of so-called active matter~\cite{marchetti2013hydrodynamics}, but also to perform mechanical work~\citep{duan2015,maggi2015micromotors}, in industrial processes~\citep{cappon2016numerical,ebbens2016active} or for biomedical and bioengineering applications~\citep{medina2015cellular,gao2014synthetic}. Many artificial microswimmers are powered by a macroscopic forcing, e.g.~using ultrasound waves~\citep{wang2012} or magnetic fields~\citep{gao2010,dreyfus2015}, or extract energy directly from their physico-chemical environment in order to self-propel at microscopic scales~\citep{moran2017phoretic,illien2017}.

Chemically-active droplets have become a prominent example of the latter. They swim using local gradients in interfacial tension created by their chemical activity, e.g. an internal chemical reaction~\cite{thutupalli2011swarming} or a solubilisation process~\cite{izri2014self}. The second category has recently received much interest and, regardless of its system-dependent physico-chemical details, involves the slow solubilisation of the droplet's fluid into the surfactant-saturated outer fluid through the formation of swollen micellar structures near the droplet's surface ~\cite{herminghaus2014interfacial,Maass16}. This effectively amounts to the emission of a chemical solute by the droplet, whose concentration near {the droplet's} surface increases its local interfacial tension~\citep{herminghaus2014interfacial}. A local excess in chemical solute thus induces a surface-driven flow and convective transport of the solute, as well as the droplet's migration away from the solute-rich region. 

{This nonlinear coupling has two fundamental consequences. First,} surface gradients of solute are enhanced by the self-induced Marangoni flows, resulting in the droplet's propulsion  when the convective transport exceeds molecular diffusion of the large micellar structures, or equivalently above a critical P\'eclet number, ${\mbox{Pe}>\mbox{Pe}_c}$~\citep{izri2014self,Michelin13b}. This instability mechanism thus underscores the fundamental role of fluid motion and convective transport in the emergence of self-propulsion.
{As a second consequence of this hydrochemical coupling}, self-propelled droplets {are repelled by their own chemical trail and are thus intrinsically negatively auto-chemotactic~\citep{Jin17,Jin19}. Like phoretic colloids, they drift down an existing gradient of chemical solute (e.g. away from other droplets) and are therefore also anti-chemotactic~\citep{Jin17,Jin19}.} 
In addition to their self-propulsion, active droplets may deform spontaneously \citep{pimienta2011,Caschera13}, exhibit chaotic behaviour~\citep{izzet2019, morozov2019nonlinear,hokmabad2020stop} and swim along curly trajectories~\citep{Suga18,Kruger16}.

Active droplets interact with and respond to ambient hydrodynamic flows or chemical gradients, generated by an external forcing, confinement or other active droplets. Chemical and hydrodynamic interactions thus influence the collective dynamics of active droplets, and alignment events~\citep{thutupalli2011swarming,moerman2017solute} and clustering~\citep{seemann2016self} have been reported in experiments. Confining boundaries likely also have a strong influence on the droplet interactions~\citep{Kruger16b,thutupalli2018flow}. 

However, the nonlinear coupling of solute transport and hydrodynamic flows distinguishes them fundamentally from phoretic swimmers and renders the  analysis and modeling of such interactions particularly challenging. Indeed, it precludes \emph{a priori} any superposition argument or rigorous dichotomy into two independent routes as typically envisioned when the chemical transport is purely diffusive~\cite{varma2019modeling,liebchen2019interactions}. Some insight  may nevertheless be gained on droplet-droplet interactions, when the large distance between the interacting droplets effectively decouples the effect of hydrodynamic and chemical signatures on the dynamics of their neighbours: a far-field estimation of hydrodynamic interactions helps therefore to understand cluster formations in the presence of boundaries~\citep{Kruger16b,thutupalli2018flow}.  Repulsive chemical interactions have also been estimated in this far-field limit by neglecting the convection of the chemical solutes, providing scaling laws that are in qualitative agreement with experimental results~\citep{moerman2017solute,meredith2019predator}. 

Yet, the droplets' self-propulsion stems fundamentally  from the convective transport of the emitted solutes. In an effort to retain such an essential ingredient, absent in far-field models, an improved modelling of active droplets{, near their self-propulsion threshold,  as moving sources of chemical solute}  identified a weak alignment {effect} of droplet collisions, and hydrodynamic interactions were shown to have little influence on such dynamics~\citep{yabunaka2016collision}. Note that a similar approach was also used recently to analyse the self-propulsion of isotropic camphor boats~\cite{boniface2019self}.

A complete modeling of the chemo-hydrodynamic interactions of active droplets was recently proposed for axisymmetric head-on collisions~\citep{lippera2020collisions}, providing for any $\mbox{Pe}$ a detailed and quantitative characterization of the role of chemical transport and hydrodynamic flow in the rebound and subsequent dynamics. Complex dynamical regimes were identified, including a delay of the droplets' rebound  at larger $\mbox{Pe}$ or the emergence of bound states of chasing droplets of different radii~\citep{lippera2020bouncing}. The subdominance of hydrodynamic interactions near the self-propulsion threshold was also confirmed~\cite{lippera2020collisions}{, as a result of the small stresslet signature of the droplets}.

These studies provide a unique insight into the detailed chemical dynamics and hydrodynamic flows. Yet, generalizing such a detailed framework to non-axisymmetric cases is technically difficult, motivating the development of reduced models to analyse the generic collisions of active droplets, that are more relevant to experimental situations. In particular, understanding the scattering or alignment properties of the droplet interactions is expected to provide critical insight on the emergence of collective dynamics in active emulsions.

To this end, the present study builds upon the observation that hydrodynamic interactions play a limited role for moderate $\mbox{Pe}$~\citep{yabunaka2016collision,lippera2020collisions}, and constructs a moving singularity model for active droplets that retains the fundamental role of convective transport in self-propulsion, neglects any hydrodynamic influence of the droplets on each other and is able to reproduce the single-droplet dynamics exactly~\citep{Michelin13b,izri2014self}, in essence improving upon a simpler version of this approach, which significantly over-estimated the self-propulsion velocity~\citep{boniface2019self}. In experiments, droplets evolve and interact within a plane, as a result of buoyancy differences~\citep{Kruger16b} or confined geometry~\citep{hokmabad2020stop,Illien2020speed}. We therefore purposely limit our analysis here to co-planar interactions of the droplets, although the model itself remains entirely three-dimensional and its application to generic 3D trajectories is straightforward.

The paper is organized as follows: the moving singularity model is first introduced  in Section~\ref{ProbForm} {and its validation for pairwise collisions of droplets, against the results of the fully-resolved interactions for head-on collisions~\citep{lippera2020collisions} is presented in Appendix~\ref{validation}}. This model is then used in Section~\ref{SymColl} to demonstrate the aligning properties of symmetric collisions. Section~\ref{GenColl} then extends the analysis to generic pairwise collisions, analysing the influence of the incoming angle and relative delay between the droplets on the emerging relative alignment and their scattering. Our findings are then summarized and discussed in Section~\ref{conclusions}, together with their implications on the collective dynamics of active droplets.

\section{Modeling droplet collisions}
\label{ProbForm}
\subsection{Description of the physical system}
We first present a general framework to analyse the collective dynamics of $N$ chemically-active droplets. Although a full description of the chemical and hydrodynamic problems is possible for any inter-droplet distance in the case of axisymmetric interactions of ${N=2}$ droplets~\cite{lippera2020collisions}, generalisation to oblique collisions or to ${N\geq 3}$ droplets quickly proves particularly challenging and intractable. Instead, we present and validate here a simplified method to obtain the droplet dynamics while still retaining the essential physical ingredients, inspired by  recent modelling of camphor boat swimmers~\cite{boniface2019self}.

Each droplet has radius $R${, density $\rho^{(i)}$} and viscosity $\eta^{(i)}$, and is immersed in a second (outer) fluid of {density $\rho^{(o)}$} and viscosity $\eta^{(o)}$. The droplets emit a chemical solute of diffusivity $D$ with a {uniform and steady flux $\mathcal A>0$ per unit area}. The solute, whose concentration field is noted $c(\bm{x})$, interacts with the droplets' surface, {thus modifying its interfacial tension. For sufficiently small concentration changes, $\gamma_c=(\partial \gamma/\partial c)$ can be considered as a constant,} so that surface chemical gradients induce local Marangoni stresses, $ {\bs\nabla_\parallel \gamma=\gamma_c \bs \nabla_\parallel c}$ \cite{anderson1989colloid}.  The resulting fluid flow advects the slowly-diffusing solute, and when ${\gamma_c>0}$ (as assumed here), this convective transport around the droplets reinforces surface concentration gradients. This mechanism is at the heart of the instability of the isotropic base state (where the droplet and fluid do not move) and of the droplets' self-propulsion beyond a minimum advection-to-diffusion ratio, or critical P\'eclet number $\mbox{Pe}_c$ \cite{izri2014self,morozov2019self}. {Note that the choice of positive $\gamma_c$ and $\mathcal{A}$ corresponds to a solute species emitted by the droplet and whose presence increases the surface tension (e.g. swollen micelles~\citep{izri2014self}). The problem remains however unchanged if both quantities are 
negative, e.g. if the solute species corresponds to surfactant monomers adsorbed by the droplet during its solubilisation~\citep{morozov2019self}.}

The droplets' activity prescribes the characteristic scale of solute concentration, ${c^*=\mathcal{A} R/D}$. We further define ${V^*=\mathcal A R \gamma_c/[D(2\eta^{(i)}+3\eta^{(o)})]}$ the characteristic solutal Marangoni drift velocity of a single droplet in a uniform concentration gradient ${\mathcal{A}/D}$~\citep{anderson1989colloid}. Using $R$, $c^*$ and $V^*$ as characteristic length, concentration and velocity scales, the  P\'eclet number, ${\mbox{Pe}=V^*R/D}$, provides a quantitative measure of the relative magnitude of convective and diffusive transport around a single droplet. {In most experimental situations,  the Reynolds and capillary numbers, $\mbox{Re}=\rho^{(o)}RV^*/\eta^{(o)}$ and $\mbox{Ca}=\eta^{(o)}V^*/R$, are both small so that inertial effects are negligible and the droplets remain spherical.}

\subsection{Self-propulsion and droplet interactions}{The $\mbox{Pe}$-dependent surface distribution of solute may introduce interfacial stresses that in turn set the flow and droplet into motion. Obtaining the droplet velocity from the surface distribution of $\gamma$ (or $c$) is a classical fluid dynamics problem.} {Indeed,} using the Reciprocal Theorem for Stokes flow, the dimensionless swimming velocity of an isolated droplet can be obtained from the chemical polarity of its surface $S_i$~\citep{masoud2019,lippera2020collisions}:
\begin{eqnarray}
\tilde{\bs v}_i=\frac{\bs v_i}{V^*}=-\frac{2}{R^2 c^*} \left\langle c \bs n \right\rangle_{S_i}.
\label{polarity}
\end{eqnarray}
where $\langle\cdot\rangle_{S_i}$ denotes the surface-averaging operator on the surface of droplet $i$ whose outward-pointing normal is $\bs n$.

By emitting solute and driving a fluid flow,  droplets influence each other both hydrodynamically and chemically. Yet, the complete modelling of two-droplet head-on collisions demonstrated that hydrodynamic interactions only have a subdominant contribution to the collision dynamics, at least for moderate $\mbox{Pe}$~\cite{lippera2020collisions}. This provides quantitative arguments for the simplified model detailed below, where direct hydrodynamic coupling between droplets is neglected, and Eq.~\eqref{polarity} is therefore valid for all the droplets. Note however that non-linear convective solute transport around each droplet, and the emergence of a chemical wake, are the essence of the self-propulsion mechanism, and should necessarily be retained at the individual droplet level, in particular their positive feedback on the polarity of the concentration distribution around the droplet.

\subsection{Moving singularity model}
To do so, the \emph{moving singularity model} proposed here describes the effect of each droplet on the concentration distribution, $c(\bs r,t)$, solely as a moving point source singularity~\cite{yabunaka2016collision,boniface2019self}, so that the chemical transport dynamics is governed by an unsteady diffusion equation:
\begin{align}
\frac{\partial c}{\partial t}&=D\nabla^2 c + 4\pi\mathcal A R^2 \sum_{i=1}^{N}\Big[\mathbf{I}+\zeta(v_i) {\bs v}_i\cdot \bm\nabla \Big]\delta(\bs r-\bs x_i(t)),
\label{unst_diff}
\end{align}
where $\bs x_i(t)$ denotes the instantaneous position of droplet $i$,  whose velocity ${{\bs v}_i=\frac{\mathrm{d} \bs r_i}{\mathrm{d} t}}$ is obtained from Eq.~\eqref{polarity}, thus accounting indirectly for the interfacial stress balance and the nonzero size of the droplet.  {Each droplet is represented as (i) a moving source of known intensity (i.e. total activity) and (ii) a moving source dipole of intensity $\zeta$ oriented along the swimming direction, that accounts empirically for the convective transport associated with near-field flows around each moving droplet. The intensity of the latter, $\zeta(v)$, depends on the velocity magnitude, and is set so that the present moving singularity model applied to a single isolated droplet matches the  exact velocity obtained from the full advection-diffusion problem for all $\mbox{Pe}$~\cite{izri2014self}. The advantage of this formulation is to retain the effect of the droplet motion on the polarity of the distribution (a chemical ``wake'' can form behind the moving point source) while allowing for linear superposition of the chemical fields created by each droplet independently.}

Note that Eqs.~\eqref{polarity} and \eqref{unst_diff} are linear so that the concentration field is the superposition of the concentration emitted by each droplet independently, ${c=\sum_jc_j}$ where $c_j$ is the solution of Eq.~\eqref{unst_diff} forced by droplet $j$ only. As a result, the velocity $\bs v_j$ of droplet $j$ is obtained from Eq.~\eqref{polarity} as the superposition of the contributions of: 
\begin{itemize}
\item{the polarity at its surface $S_j$ of its own concentration footprint, i.e. the asymmetry of its {chemical }wake, ${\bs\Pi_j=-\langle c_j\bs n\rangle_{S_j}}$,}
\item{the polarity at its surface $S_j$ of the concentration emitted by other droplets,  ${-\langle c_k\bs n\rangle_{S_j}}$ with ${k\neq j}$.}
\end{itemize}

In the case of a single isolated droplet  in steady self-propulsion with velocity ${\bs v}$, the dimensionless concentration field writes:
\begin{align}
\frac{c}{c^*}&=\left(\mathbf{I}+\zeta( v) {\bs v}\cdot\nabla\right)\cdot\left\{\frac{R}{r}\mathrm{exp}\left[-\frac{\left({v}r+{\bs v}\cdot\bs r\right)}{2D}\right]\right\}\nonumber\\
&=\frac{R}{r}\left[1-\zeta( v)\left(\frac{{\bs v}\cdot\bs r}{r^2}+\frac{{v}({v}r+{\bs v}\cdot\bs r)}{2Dr}\right)\right]\nonumber\\
&\qquad\qquad\times\mathrm{exp}\left[-\frac{\left({v}r+{\bs v}\cdot\bs r\right)}{2D}\right],
\label{sourcedip}
\end{align}
where $\bs r$ is the radial vector taken from the droplet's center. Using this result in Eq.~\eqref{polarity}, and defining {${\lambda=vR/2D=(\tilde v\mbox{Pe})/2}$}, provides the dipole intensity $\zeta$ uniquely in terms of the exact result for the non-dimensional velocity $\tilde{v}_\textrm{sp}(\mbox{Pe})$\citep{izri2014self}:
\begin{align}
\frac{\zeta(\lambda)}{R^2/D}=\frac{1}{2}\left[\frac{\lambda^2e^{\lambda}\tilde{v}_\textrm{sp}/2-\lambda\cosh\lambda+\sinh\lambda}{2\sinh\lambda-(2\lambda+2\lambda^2+\lambda^3)e^{-\lambda}}\right]\cdot\label{dipoleint}
\end{align}
Note that in its non-dimensional form and for a fixed viscosity ratio (which will affect $\tilde{v}_\textrm{sp}(\lambda)$), the dipole intensity only depends on $\lambda$. In the following, ${\eta^{(i)}=\eta^{(o)}}$ is assumed.

Equations~\eqref{polarity}, \eqref{unst_diff} and \eqref{dipoleint} together with the definition of the droplets' velocity ${\bs v_i=\frac{\mathrm{d}\bs r_i}{\mathrm{d}t}}$ provide a closed set of equations for the droplets' dynamics and concentration distribution. These equations are solved spectrally in a large periodic domain (see Appendix~\ref{fourier}). 

In the following, we apply this moving singularity model to analyse {the }collision dynamics of ${N=2}$ droplets. Initially, the droplets are located at a distance $d$ that is large enough that they essentially behave as isolated and have a steady self-propulsion velocity of magnitude $v_\textrm{sp}(\mbox{Pe})$. After the encounter with the second droplet, each droplet recovers a steady self-propulsion regime albeit with a modified orientation.
\begin{figure}
\begin{center}
\includegraphics[width=.48\textwidth]{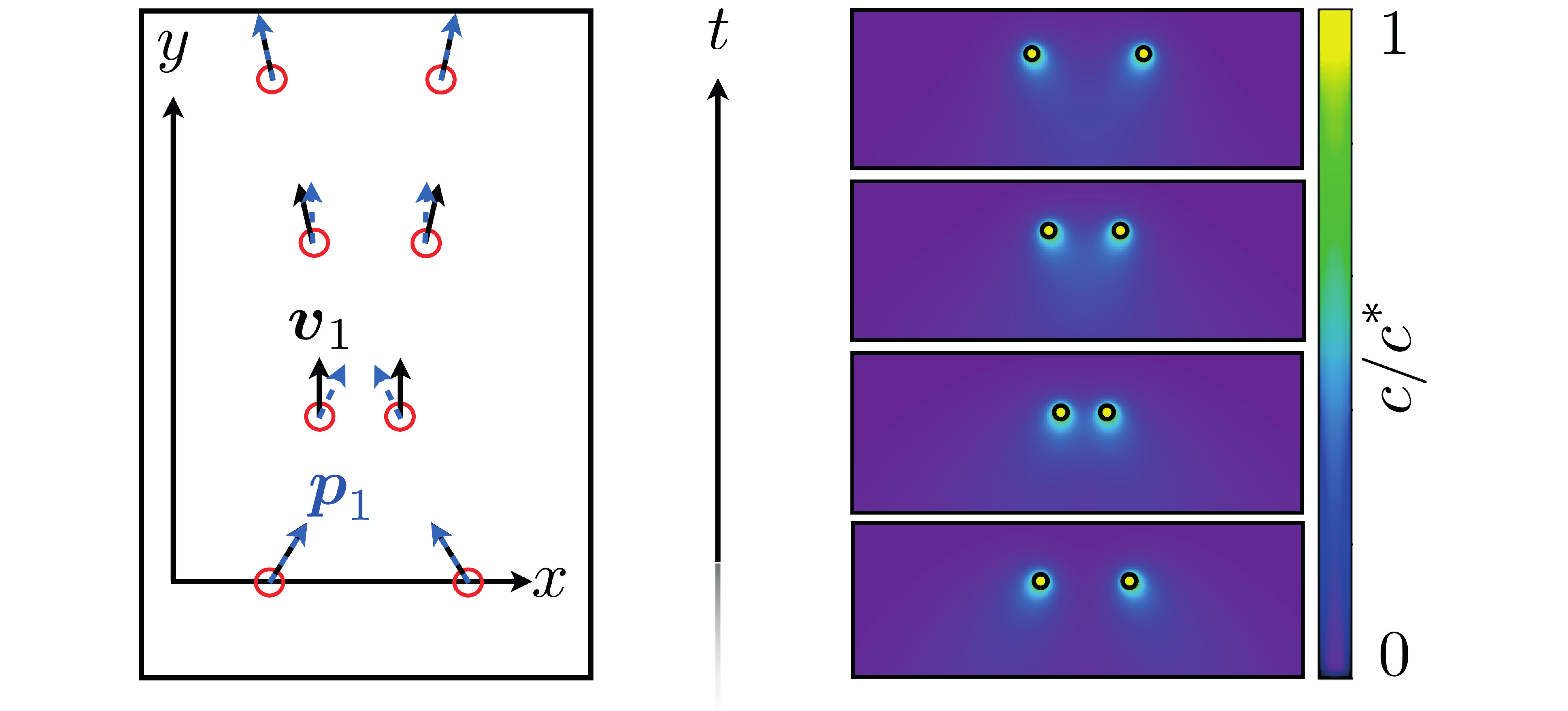}
\caption{(Left) Oblique symmetric collision of two active droplets in exact collision course. Schematic snapshots of the polarity direction ${\bs p_1=\bs \Pi_1/\Pi_1}$ and velocity $\bs v_1$ are provided along the collision (left) as well as the corresponding concentration fields (right) for ${\dircos_0=0.5}$ and $\mbox{Pe}=6$. }
\label{snapshots_collision}
\end{center}
\end{figure}
In order to validate the relevance and accuracy of the present numerical model, its predictions for the head-on collision of two droplets were compared to the exact results~\citep{lippera2020collisions} (see Appendix~\ref{validation}).

Note that the model proposed here is fully three-dimensional. Yet, motivated in part by the quasi-2D motion of active droplets in experiments, we restrict our discussion of the collision problem to planar trajectories of both droplets (the chemical dynamics remains however three-dimensional). {The sensitivity of the results to 3D misalignment of the droplets is analysed in Appendix~\ref{app:3d}.}

\subsection{Oblique collisions}
Active swimming droplets are anti-chemotactic, and thus swim away from the zones of higher concentration, e.g. their own {chemical} wake or the proximity with other emitting droplets~\cite{izri2014self,Maass16,Jin17}. During the encounter of two droplets, the confinement-induced accumulation of the emitted solute between them modifies the orientation of their chemical polarity and velocity: after a transient interaction, the droplets swim away from each other in different and modified directions (Figure~\ref{snapshots_collision}). 

In a head-on collision, the axisymmetry of the problem imposes that the droplets' velocity is strictly zero when they are closest to each other~\citep{lippera2020collisions}. This is however not necessarily the case in oblique collisions for which the droplets can maintain a non-zero velocity at all time, and the chemical wake can rotate around the droplet as a result of the change in swimming direction (see figure~\ref{snapshots_collision}). The outcome of such oblique collisions is therefore not obvious, in particular for the final direction of the droplets as they swim away from each other, and is intimately linked to the detailed unsteady dynamics of the droplets' chemical wake. 
\begin{figure}[t]
\centering
\includegraphics[width=0.5\textwidth]{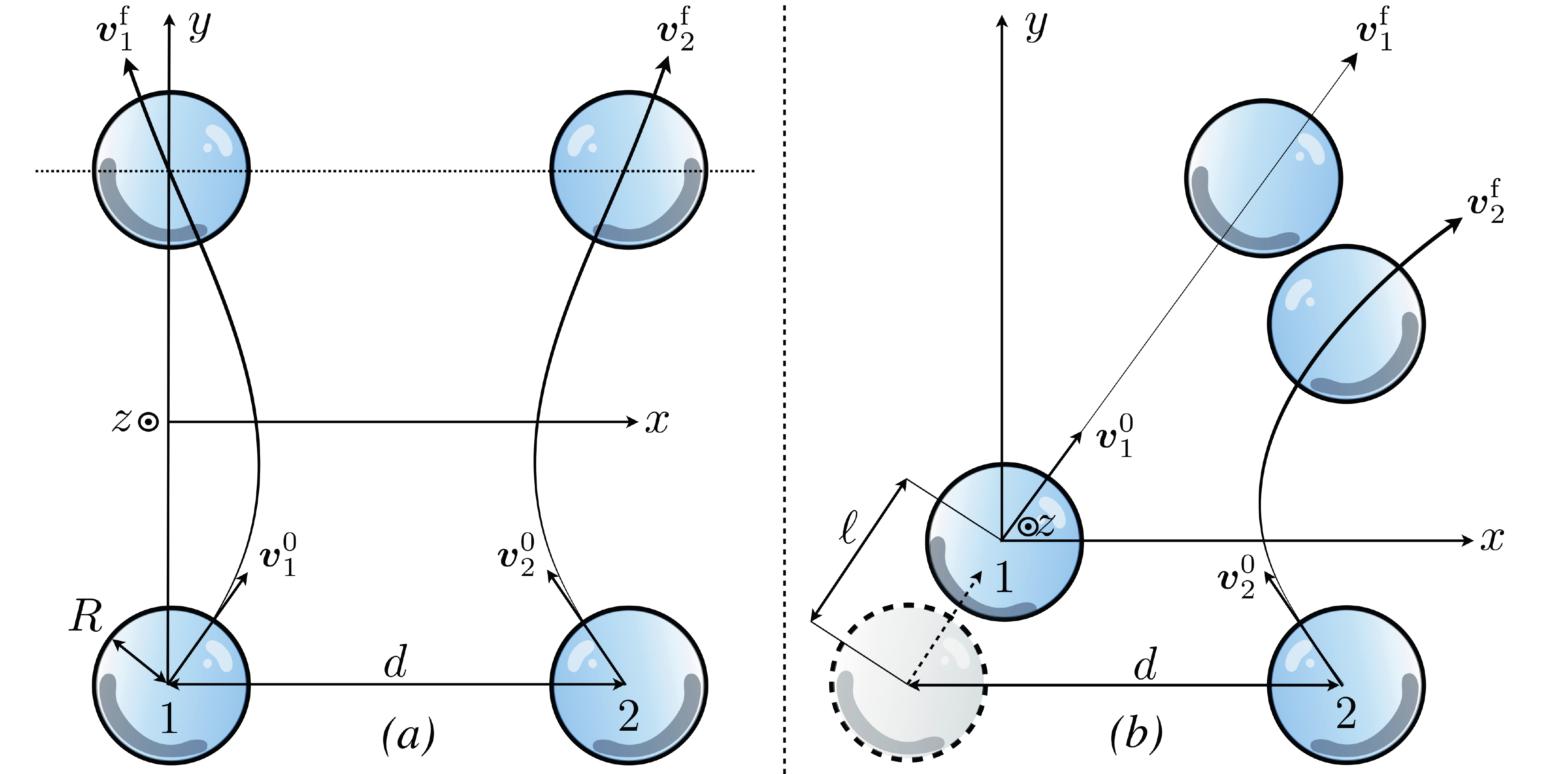}
\caption{(a) Symmetric and (b) generic {co-planar} collisions of two active droplets. In $(a)$, the two droplets are exactly on a collision course and the problem is symmetric.  For generic collisions $(b)$, $\ell$ denotes the lead distance of droplet $1$ on the second droplet{, and the} final velocity directions of droplets~1 and 2 are not symmetric anymore.}
\label{SchemaCollisions}
\end{figure}
 
 This is the main focus of the present paper and in the following, we analyse in detail the influence of generic droplet-droplet interactions on their directional dynamics. Symmetric oblique collisions are first analysed, where both droplets are initially exactly on a collision course (Section~\ref{SymColl}, Figure~\ref{SchemaCollisions}a). By this terminology, we mean that the droplets are initially at the same distance of the crossing point of their incoming trajectories: the problem  maintains therefore a reflection symmetry at all times. In a second step, the general case is considered, where one of the droplets (termed droplet 2 by convention) is lagging by a finite distance  (Section~\ref{GenColl}, Figure~\ref{SchemaCollisions}b).

\section{Symmetric oblique collisions}
\label{SymColl}

\subsection{Collision-induced alignment}
We first consider the symmetric collision of two active droplets, initially separated by a large distance ${d\gg R}$ and swimming towards each other (Figure~\ref{SchemaCollisions}a). The droplets' motion is completely symmetric and we thus focus exclusively on the dynamics of the left-most droplet (droplet 1). 
When the two droplets are sufficiently far from each other, the solute concentration emitted by each of them does not influence the other's swimming motion. As a result, long before and after the collision, each droplet swims as if it was isolated, with a constant velocity $v_\textrm{sp}$ and along a straight trajectory.

\begin{figure}
\centering
\includegraphics[width=0.5\textwidth]{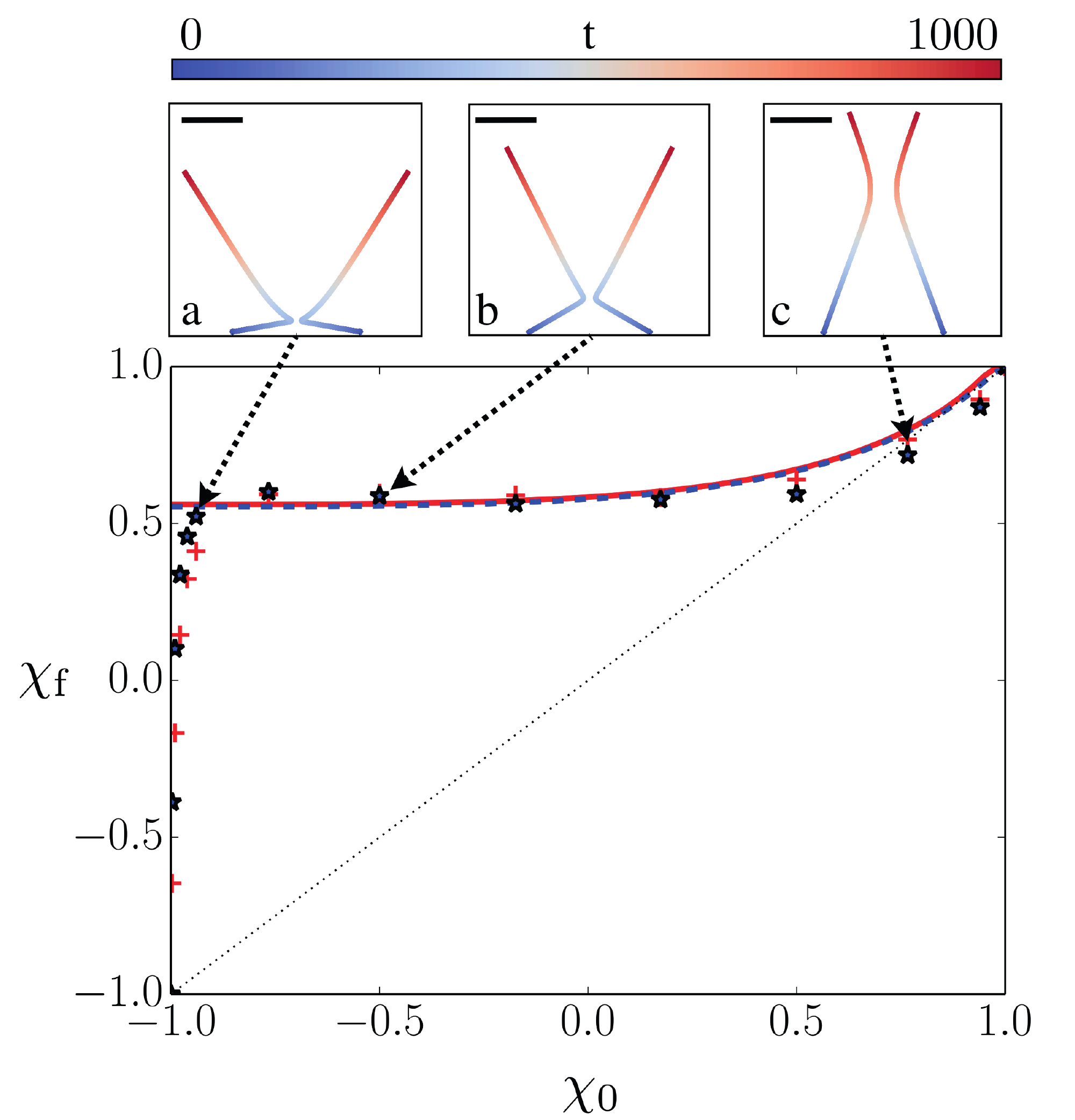}
\caption{Reorientation of the droplets' dynamics in a symmetric collision for ${\mbox{Pe}=6}$ (red) and ${\mbox{Pe}=8}$ (blue). The cosine of the velocity {relative} orientation is compared in the initial and final steady propulsion regimes when the droplets are far away from each other. The results of the moving singularity model (symbols) are compared to the prediction of the reduced model {of Section~\ref{sec:minim_model} }(lines). {To illustrate the alignment dynamics, the droplets' trajectories are represented for three representative cases in the top panels: nearly-head-on collision ($\chi_0=-0.94$), strong alignment ($\chi_0=-0.5$) and symmetric rebound ($\chi_0=0.77$).}}
\label{scattering_angle}
\end{figure}
Defining  ${(\bs v_1^0,\bs v_2^0)}$ and ${(\bs v_1^{\text{f}},\bs v_2^{\text{f}})}$ the initial and final droplets' velocities (Figure~\ref{SchemaCollisions}a), and their relative direction cosines and sines,
\begin{align}
\dircos&=\cos(\bs v_1,\bs v_2)=\frac{\bs v_1\cdot\bs v_2}{|\bs v_1||\bs v_2|}, \label{dircos}\\
\dirsin&=\sin(\bs v_1,\bs v_2)=\frac{\bs e_z\cdot(\bs v_1\times\bs v_2)}{|\bs v_1||\bs v_2|},\label{dirsin}
\end{align}
the effect of the collision on the droplets' alignment can be quantified by relating their relative direction cosine before ($\dircos_0$) and after ($\dircos_f$) the collision (Figure~\ref{scattering_angle}).

In a sharp contrast with a perfect elastic {collision} of rigid passive spheres (for which ${\dircos_0=\dircos_f}$), the symmetric collision of active droplets results in a systematic alignment of the droplets regardless of their initial relative {direction} (${\dircos_f>\dircos_0}$). This alignment is most striking for rather frontal collisions for which ${\dircos_0\in]-1,0]}$ which corresponds to droplets initially heading mostly toward each other. Qualitatively, we understand this as the result of the droplets coming closer to each other in such configurations (smaller $d_\textrm{min}$): the chemical repulsion induced by the other droplet, at the origin of the droplet's reorientation and rebound, is a decreasing function of their relative distance.

A second distinctive feature of Figure~\ref{scattering_angle} is the emergence of a plateau for ${\dircos_0\in[-0.9,0.5]}$: within that range, the droplets swim away from each other with a relative {direction cosine} ${\dircos_f\approx 0.5}$ that is essentially independent of their incoming orientation. For greater $\dircos_0$ (i.e. almost parallel incoming trajectories), the elastic rebound dynamics is recovered, ${\dircos_f\approx\dircos_0}$, as a result of the weak interaction of the droplets which remain far from each other at all times. For {almost head-on collisions (${\dircos_0<-0.9}$)}, the final direction of the droplets is extremely sensitive to the exact impinging angle. Perfect head-on collisions (${\dircos_0=-1}$) result in a normal rebound (${\dircos_f=-1}$) by complete reversal of the chemical wake and of their swimming velocity; but a small departure from this situation (e.g. ${\dircos_0=-0.93}$) results in a sharp alignment of the droplets (${\dircos_f\approx 0.5}$). This sensitivity is intimately linked to the complex {reorganisation} of the chemical polarity in this type of collisions, and suggests furthermore that purely head-on collisions are unstable.

\begin{figure}
\centering
\includegraphics[width=0.5\textwidth]{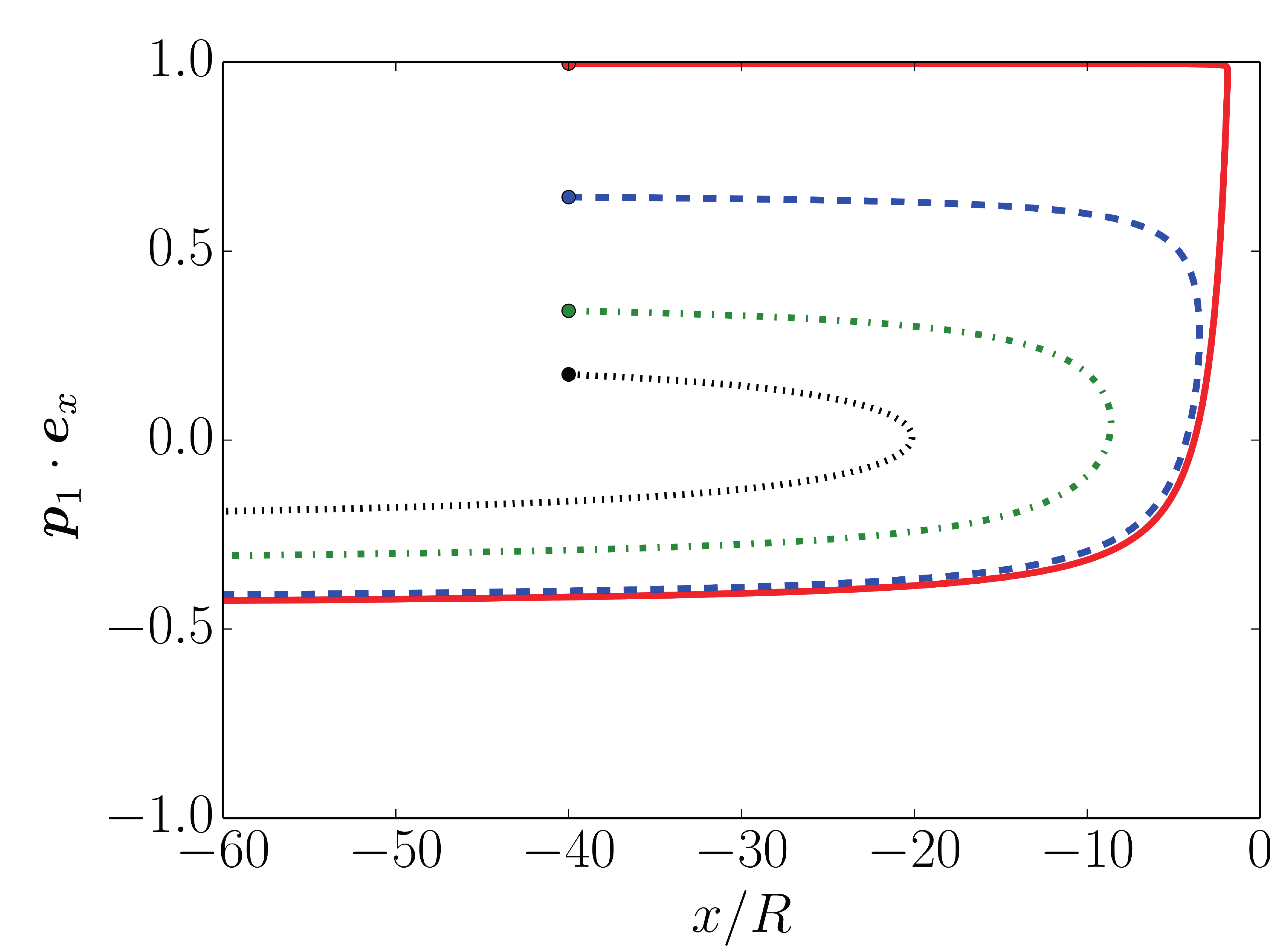}
\caption{{Phase-space representation of the collision in terms of droplet 1's horizontal position $x$ and its polarity direction $\bs p_1\cdot \bs e_x$} for different initial angles: solid red line ${\dircos_0=-0.98}$, dashed blue line ${\dircos_0=0.17}$, dashed dotted green line ${\dircos_0=0.77}$ and black dotted line ${\dircos_0=0.94}$.}
\label{schematic_c}
\end{figure}

Greater physical insight in the collision dynamics is provided by the dynamic evolution of the chemical wake  which is represented at different stages of a collision with {${\dircos_0=0.5}$} and ${\mbox{Pe}=6}$ on Figure~\ref{snapshots_collision}. The {finite-time diffusion of solute} is retained in the moving singularity model, and it should thus be noted that{, as a result,} the direction of the wake created by a droplet's own chemical footprint $\bs p_i$ does not align instantaneously to its velocity $\bs v_i$, but instead takes a finite time to adjust to changes in the swimming direction induced by the additional drift created by the other droplet's chemical footprint. 

\subsection{Minimal collision model}\label{sec:minim_model}
 These observations provide the basic ingredients of an even simpler dynamic model for the collision, which is referred to in the following as \emph{minimal collision model}, with only two degrees of freedom: the separation distance $d$ between the droplets and the direction, $\bs p_1$, of the chemical wake of the left-most droplet, i.e. the polarity of its own concentration footprint $c_1$ at its surface $S_1$. When the droplet is isolated, its velocity is aligned with $\bs p_1$, i.e. ${\bs v_1={\frac{2V^*}{R^2c^*}}\bs\Pi_1=v_\textrm{sp}\bs p_1}$. In a more general situation, invoking \eqref{polarity} and \eqref{unst_diff}, its total velocity $\bs v_1$ is obtained  as the sum of two contributions,
\begin{align}
\label{veq}
\bs v_1={\frac{2V^*}{R^2c^*}}\bs \Pi_1 +\bs v^r_1,
\end{align}
with $\bs v_1^r$ the chemical repulsive drift induced on droplet 1 by the chemical footprint $c_2$ of droplet 2. For simplicity, this repulsion velocity $\bs v^r_1$ is modelled here as resulting from the chemical gradient generated by a fixed source of intensity ${4\pi\mathcal A R^2}$ at the instantaneous location of the second droplet, thus retaining only the slowest decaying signature of the moving singularity model: 
\begin{align}
\label{vr}
\bs v_1^r=M\frac{\bs x_1-\bs x_2}{|\bs x_1-\bs x_2|^3}=-\frac{M}{d^2}\bs e_x,
\end{align}
where $M$ is a positive constant characterizing the mobility of a passive droplet in a chemical gradient. In the full moving singularity model, the evolution of a droplet's {chemical }wake in response to changes in its total velocity is a complex process and involves both changes of direction and magnitude in the chemical self-polarity (i.e. corresponding to its own chemical footprint) under the effect of diffusion and of the droplet's translation. The simplified model considered here is based on two main physical features of that process, namely that  the polarity (i) evolves  in response to changes in the droplet's velocity and (ii) relaxes with a finite delay $\tau$ to the droplet's swimming direction in steady state. As a result, and further assuming that the magnitude of the self-polarity $\Pi_1$ does not change in time, the evolution equations for the velocity $\bs v_1$ and {chemical }wake direction $\bs p_1$ become (see Appendix~\ref{reducedmodel})
\begin{align}
\bs v_1&=v_\textrm{sp}\bs p_1+\bs v_1^r,\label{simp_vel}\\
\frac{d \bs p_1}{dt}&= \frac{(\bs p_1 \times \bs v_1)\times \bs p_1}{\tau v_\textrm{sp}}= \frac{\kappa(\bs p_1 \cdot \bs e_y)(\bs e_z \times \bs p_1)}{d^2} ,
\label{polarity_torque}
\end{align}
where ${\kappa=M/(\tau v_\textrm{sp})}$ is a positive constant.
The reorientation of the polarity is then solely the result of the chemical repulsion by the other droplet, Eq.~\eqref{polarity_torque}. 

The essence of the collision dynamics observed for the full system is well captured by this simplified model: as the droplets get closer to each other, the chemical repulsion reduces the magnitude of their relative velocity (i.e. the component along the $x$-axis on Figure~\ref{snapshots_collision}) which eventually vanishes at a distance $d_{\text{min}}$; at that instant, the component of self-propulsion in the $x$-direction  balances the chemical repulsion exactly. However, $\bs p_1$ is not yet aligned with $\bs v_1$ as a result of the finite time delay $\tau$: the {chemical }wake continues to rotate for a finite time reducing (then reversing) the self-propulsion component along $\bs e_x$ which can not balance the chemical repulsion ${\bs v_1^r\parallel -\bs e_x}$: this generates the rebound of droplet~1  away from its neighbour.

The results of the minimal collision model can now be compared with those of the original dynamics obtained from the moving singularity description. In the simplified model, the self-propulsion velocity $v_\textrm{sp}$ is directly imposed by the choice of P\'eclet number \citep{izri2014self}; for fixed $\mbox{Pe}$, the minimal model, Eqs.~\eqref{simp_vel}--\eqref{polarity_torque}, therefore includes a single fitting parameter $\kappa$. Figure~\ref{scattering_angle} {compares} the final relative {direction} characterised by {$\dircos_f$} predicted by the simplified model (solid lines) as a function of {$\dircos_0$} with the complete numerical results obtained for ${\mbox{Pe}=6}$ and ${\mbox{Pe}=8}$ (respectively red crosses and blue stars).  We note that for ${\dircos_0>-0.93}$ the simplified model captures the emergence of the {constant $\dircos_f$ plateau} for a large range of approaching angles.

 In head-on collisions, the norm of the droplets' polarity vanishes to zero at the moment they are closest (see figure \ref{Headon}). The \emph{minimal collision model} only describes the direction of polarity and not its magnitude, and is therefore intrinsically unable to reproduce the physics of such specific configurations. 

The problem is invariant by translation along $y$, and the simplified model provides a two-degree-of-freedom description of the collision dynamics, namely the $x$ position of droplet 1 and the angle of its polarity direction $\bs p_1$ with the $y$-axis. The dynamics can therefore be fully {characterised} by the system trajectories in the ${(x,\bs p_1\cdot \bs e_x)}$-plane, Figure~\ref{schematic_c}. We note an accumulation of the trajectories near the minimum distance and onto the trajectory emerging from a perturbation of the head-on collision (${\chi_0\approx -1}$), which is indeed consistent with the emergence of the plateau-behaviour of the outgoing {relative angle}, regardless of the initial orientation of the droplets.

\section{Generic oblique collisions}
\label{GenColl}

We now turn to the general problem of asymmetric or delayed collisions, which are characterized below using the full moving singularity model introduced in Section~\ref{ProbForm}.  In such collisions, droplet~$2$ is initially located further than droplet~$1$ by a ``delay'' distance $\ell$ from the virtual crossing point of the initial trajectories (Figure~\ref{SchemaCollisions}b). In contrast with many active particle systems, active droplets leave a chemical ``trail'' that extends over several tens of radii and is known to influence critically their collective dynamics and trajectories~\cite{Jin18,Jin19}: when crossing another droplet's trail, a second droplet is expected to be deviated away or repelled by the slowly-diffusing solute left by the first droplet when it went by. This interaction and deviation is obviously strongest for close interactions, i.e. when $\ell$ is small. 

In the following, we analyse the possible outcome of such general encounters of two droplets and the impact on their subsequent relative dynamics. By convention, and without any loss of generality, droplet~$1$ (resp. $2$) is initially located on the left (resp. right) and both droplets are heading toward each other, so that $\dirsin>0$, see Eq.~\eqref{dirsin}. Depending both on their initial relative alignment, $\dircos_0$, and the delay length $\ell>0$ of droplet 2, the droplets can either \emph{cross paths} (${\dirsin_0\dirsin_f>0}$) or \emph{rebound} (${\dirsin_0\dirsin_f<0}$).

\begin{figure*}[ht]
\begin{center}
\includegraphics[width=\textwidth]{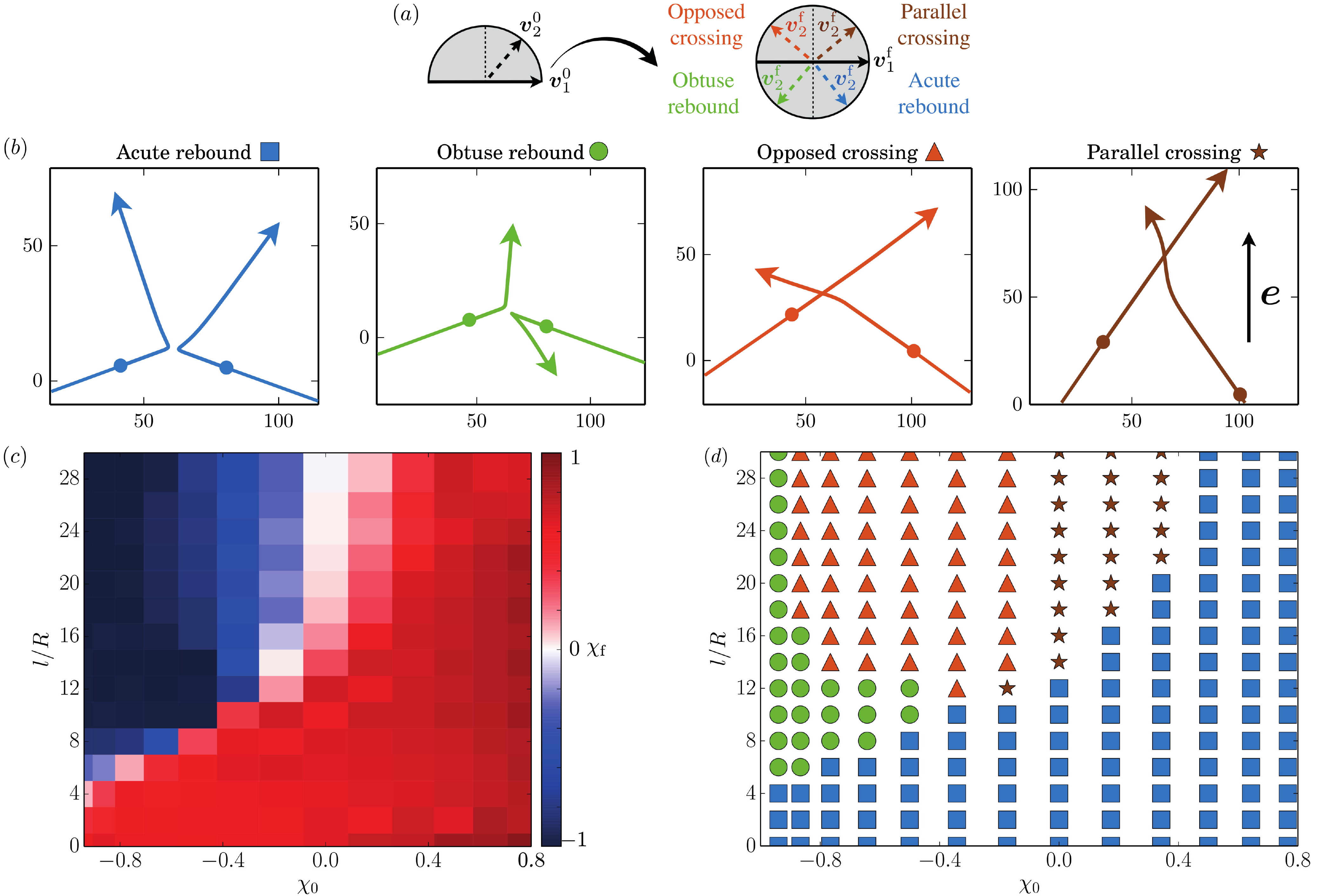}
  \caption{Four possible regimes following  asymmetric (delayed) collisions of two droplets. $(a)$: Initial and final relative orientations of the droplets in each regime. $(b)$: Illustration of each regime: acute rebound (blue, ${\dircos_0=-0.77}$, ${\ell=2}$), obtuse rebound (green, ${\dircos_0=-0.34}$, ${\ell=30}$), opposed crossing (red, ${\dircos_0=-0.77}$, ${\ell=8}$)  and parallel crossing (brown, ${\dircos_0=0.34}$, ${\ell=30}$). In each case, the droplets' trajectories are provided together with their position at a given time before the collision. $(c)$: Final alignment, $\dircos_f$, as a function of the initial alignment $\dircos_0$ and delay length $\ell$. $(d)$: Phase diagram of the collision outcome depending on the initial relative alignment, $\dircos_0$, and delay length, $\ell$. }
\label{Phasediag}
\end{center}
\end{figure*}

Each of these two general behaviours is further divided into two different regimes depending on the sign of their final relative alignment, $\dircos_f$ (Figure~\ref{Phasediag}a): 
\begin{itemize}
\item{In the \emph{crossing regimes}, ${\dirsin_f<0}$, droplet $2$ passes through the chemical wake of droplet~$1$. In their final state, the droplets can either swim in opposite directions (\emph{opposed crossings}, ${\dircos_f<0}$, red color on Fig.~\ref{Phasediag}), or in the same direction (\emph{aligned crossings}, ${\dircos_f>0}$, brown color on Fig.~\ref{Phasediag}).} 
\item{In the \emph{rebound regimes}, ${\dirsin_f>0}$, droplet~$2$ is repelled by droplet~$1$ and its chemical trail, and is deviated away before crossing its path (for sufficiently small $\ell$, droplet~$1$ may also be deviated by the oncoming droplet~$2$). Again, the final relative orientation of the droplets provides a distinction between \emph{acute rebounds} ($\dircos_f>0$, blue color on Fig.~\ref{Phasediag}) and \emph{obtuse rebounds} (${\dircos_f<0}$, green color on Fig.~\ref{Phasediag}).}
\end{itemize}
These four different regimes are illustrated on Fig.~\ref{Phasediag}(b) and the influence  on their selection of the delay length, $\ell$, and initial alignment, $\dircos_0$, is fully characterised below. 

When $\ell$ is small, the problem is almost symmetric so that both droplets rebound around the same time under the effect of their repulsive interaction, oriented orthogonally to their average initial direction, $\bs e$. This symmetry is broken when $\ell$ is increased; as a result, when the droplets are closest, the repulsive interactions experienced by each of them point along distinct directions, and can  lead to completely different dynamics for the leading and trailing droplets.

Section~\ref{SymColl} demonstrated that symmetric collisions (${\ell=0}$) systematically lead to acute rebounds provided ${\dircos_0>-0.98}$, while obtuse rebound are observed for strictly head-on collisions (${\dircos_0= -1}$). It is therefore no surprise that  such observations are maintained for small enough delay length $\ell$\footnote[2]{Note that Figs.~\ref{Phasediag}c-d do not include strict head-on and parallel collisions, ${\dircos_0=\pm 1}$ for which $\ell$ can not be defined.}. 

In fact, acute rebounds are observed for most initial relative orientation when ${\ell/R<5}$, and for even larger delay lengths when the droplets are initially swimming in rather parallel directions (${\dircos_0>0}$). For more frontal collisions (${\dircos_0<-0.3}$), alignment of the droplets and acute rebounds are still observed for small $\ell$, but the second (delayed) droplet follows a drastically different dynamics above a critical delay length $\ell_c$, leading to obtuse rebounds. To understand this acute-to-obtuse rebound transition,  the detailed dynamics of the droplets must be analysed in the \emph{interaction region}, loosely defined here as the region where their relative distance is minimum.

We noted previously the asymmetry of the chemical footprint of a swimming droplet: most of the chemical released by the droplet is left in its wake. As a result, the interaction region  is almost solute-free as the first (leading) droplet crosses it, and for larger delay length, droplet 1 is therefore only weakly deviated. In contrast, when it finally crosses the interaction region, droplet 2 is repelled by the {chemical }wake of droplet 1 in a direction that depends both on $\ell$ (i.e. how long ago droplet 1 went by) and $\dircos_0$. For large enough $\ell$ and small $\dircos_0$ (droplets heading toward each other), this repulsion includes a component along $-\bs e$. This justifies the existence of a critical delay length $\ell_c(\dircos_0)$ for the acute-to-obtuse rebound transition observed on Fig.~\ref{Phasediag}(d): for ${\ell\approx\ell_c}$, the repulsion of droplet 2 along $-\bs e$  compensates its initial velocity component along $+\bs e$, which increases with $\dircos_0$. As the interaction strength decreases with the droplet separation, $\ell_c$ is an increasing function of $\dircos_0$, which is consistent with the positive slope of the separation between acute and obtuse rebound regimes on Fig.~\ref{Phasediag}(d).

For larger $\ell$ (typically ${\ell\gtrsim 5-10}$), the sign of the droplets' alignment, $\dircos$, is conserved between the initial and final configurations: droplets initially swimming along rather parallel directions (${\dircos_0>0}$) experience an \emph{acute rebound}  or a \emph{parallel crossing} while droplets heading more directly toward each other ($\dircos_0<0$), experience an \emph{obtuse rebound} or an \emph{opposed crossing}. In both cases, a rebound-to-crossing transition is observed when the delay $\ell$ is large enough (Figure~\ref{Phasediag}d). This is consistent with the physical intuition that the droplets essentially do not interact and maintain a straight trajectory for sufficiently large $\ell$, as the {solute emitted by }the leading droplet has diffused away by the time the second droplet crosses the interaction region. 

We note that the second critical delay length ${\ell_c^*(\dircos_0)}$ required for this rebound-to-crossing transition varies non-monotonically with $\dircos_0$: it is minimum for ${\dircos_0\approx 0}$ and diverges for rather parallel or head-on configurations (${\dircos_0\rightarrow\pm 1}$). 
This feature results mainly from the non-trivial variations of the minimum distance of the two droplets with $\ell$ and $\dircos_0$ as discussed below.

The minimum distance reached by the droplets generally increases with $\dircos_0$ (see Fig.~\ref{schematic_c} for the case of symmetric collisions, $\ell=0$), and diverges for parallel configurations (${\dircos_0\rightarrow 1}$). To experience a rebound, the droplets must reverse the component of their relative velocity normal to $\bs e$, which is proportional to $\sqrt{1-\dircos_0}$. For greater initial alignment (larger $\dircos_0$), this is achieved at greater distances (the chemical repulsion decreases as $1/d^2$). Furthermore, the interaction of the droplets is stronger as their alignment increases due to the angular asymmetry of their chemical wake (see Eq.~\eqref{sourcedip}). As a result, a much greater delay length $\ell$ is required to avoid a rebound when the droplets swim initially parallel to each other, which is consistent with the critical delay $\ell^*_c$ for a rebound-to-crossing transition being an increasing function of $\dircos_0$ when ${\dircos_0>0}$ (Figure~\ref{Phasediag}d).

Additionally, a first estimate of the minimum distance of the two droplets is provided by the minimum distance reached by two non interacting droplets ${d_\textrm{min}^*\sim\ell\sqrt{1+\dircos_0}}$, which is always small for head-on configurations, even for large $\ell$. As a result, a rebound is observed for larger delay $\ell$ when ${\dircos_0\rightarrow -1}$ (head-on collisions), which is consistent with $\ell_c^*$ being a decreasing function of $\dircos_0$ for ${\dircos_0<0}$ (Figure~\ref{Phasediag}d).

Finally, in addition to the phase diagram of Figure~\ref{Phasediag}(d), Figure~\ref{Phasediag}(c) provides the evolution of the final relative alignment of the droplets, $\dircos_f$, in the ${(\dircos_0,\ell)}$-parameter space. Two regions can be distinguished on this figure. Acute rebounds (small $\ell$ or large $\dircos_0$) are characterised, as for symmetric collisions, by a rather fixed directional outcome which {corresponds} to a general alignment of the droplets ($\dircos_f$ is almost constant and greater than $\dircos_0$). This region is separated by a sharp transition from the rest of the map in which the relative direction is mostly conserved (${\dircos_f\approx\dircos_0}$) and depends only weakly on $\ell$. This sharp transition stems from sudden changes in the reorientation direction of the trailing droplet under the effect of the chemical wake left behind the leading droplet. It emphasizes the sensitivity of the collision outcome to $\ell$ and the scattering ability of such collisions on  the collective behaviour of the droplets.

\section{Conclusions}\label{conclusions}
Swimming droplets influence each other's dynamics through the wake of chemical solute  they generate in order to self-propel. These chemical interactions are repulsive and  have been identified as the dominant contribution to the droplets' collective dynamics, both in experiments~\cite{Jin17} and from a complete modelling of the two-droplet dynamics~\cite{lippera2020collisions,lippera2020bouncing}. Based on this observation, this work proposes a general simplified framework in terms of moving singularities to analyse the collisions of $N$-droplet collisions. Building upon a detailed understanding of the axisymmetric configuration, for which a full solution of the chemo-hydrodynamic system is available, this model is then exploited to characterise in detail the generic (oblique) planar collisions of two droplets, which is more relevant to experimental conditions. 
In fact, it is shown that purely axisymmetric or head-on collisions are very specific in terms of the droplets' wake dynamics whose {chemical }polarity must vanish in a rebound, while it is able to rotate around the droplet in a more generic setting. As a result, mostly (but not strictly) head-on collisions lead to a significant scattering.

Our results show that symmetric collisions systematically align the droplets (${\dircos_f\geq \dircos_0}$), and lead to a surprisingly constant relative final alignment $\dircos_f$, regardless of the incoming orientation $\dircos_0$. This phenomenon was proved to result essentially from the reorientation dynamics of each droplets' own wake during the collision, and was rationalised using a simple two-degree-of-freedom model in terms of the {chemical }wake orientation and inter-droplet distance. 

This alignment ability of the droplet interactions is maintained for significant asymmetry in the droplets' oncoming dynamics, at least for effective delay length of a few droplet radii. When the asymmetry of the droplet interaction is greater (i.e. when the trailing droplet crosses the interaction region long enough after the leading droplet did), the interaction outcomes are much more diverse, and both rebound regimes (where the droplets' relative velocity is reversed) and crossing regimes (where the droplets are only deviated away from their original trajectory) were observed. Two sharp transitions between fundamentally different outcomes were observed as a result of the  strong sensitivity of the trailing droplet's trajectory and final heading to the exact timing of its crossing of the solute-rich region left behind the first droplet.  

The alignment ability of the droplet interactions is expected to favour a certain collective coherence of the droplets' trajectory at large scales. In contrast, sharp transitions in the final dynamic regime denote the scattering ability of such collisions, which provide the droplets with the ability to explore new spatial directions.

For simplicity, we specifically focused here on planar dynamics of the droplets, still accounting for the fully-3D diffusion of the chemical solute. {A preliminary analysis of more generic (3D) trajectories (Appendix~\ref{app:3d}) further suggested that our findings are robust with respect to three-dimensional perturbations.} The {detailed} stability of such planar collisions remains however to be studied. Yet, in experimental situations, an external mechanism (in general confinement or buoyancy-induced trapping close to a boundary) maintains such planar configurations and we therefore expect the present findings to be relevant to such situations. Note also, that the present framework {can} be generalised to include the effect of confinement on the solute dynamics, e.g. exploiting the linearity of the chemical problem through the method of images, {in order to analyze the effect of confinement on the collective dynamics of active droplets~\citep{Kruger16b,thutupalli2018flow}.}

{Finally, the present model purposely considers a simplified physico-chemical description of the system in order to focus on the effect of finite-time diffusion of the solute on the droplets' interactions. Although this is beyond the scope of the present work, it should be noted that this model can easily be generalised to account for multiple active species or more complex kinetics, e.g. to analyse the effect of different solute diffusivities or the activity inhibition by the local micellar content of the solution.}
\section*{Conflicts of interest}
There are no conflicts to declare.

\section*{Acknowledgements}
This project has received funding from the European Research Council (ERC) under the European Union's Horizon 2020 research and innovation programme (grant agreement No 714027 to SM).
\appendix

\section{Numerical solution of the moving singularity model}
\label{fourier}
The dynamics of $N$ droplets is obtained by solving Eqs.~\eqref{polarity}, \eqref{unst_diff} and \eqref{dipoleint} in a large periodic domain of dimensions ${(L_x, L_y, L_z)}$,  using a Fourier decomposition of the concentration field:
\begin{equation}
c(\bs r,t)=\sum_{l,m,n}\hat{c}_{l,m,n}(t)e^{2\pi i\bs k\cdot \bs r}\quad,
\label{Fourierdeco}
\end{equation}
where ${\bs k=(l/L_x)\bs e_x+(m/L_y)\bs e_y+(n/L_z)\bs e_z}$. The velocity of droplet $i$ is then computed by substituting Eq.~\eqref{Fourierdeco} into Eq.~\eqref{polarity}, as
\begin{equation}
\bs v_i=\frac{V^*}{ c^* }\sum_{l,m,n}\frac{\hat{c}_{l,m,n}e^{2\pi i \bs k \cdot \bs x_i}}{\pi R k^2}\Big(\cos(2\pi R k)-\text{sinc}(2\pi Rk)\Big)\bs k.
\end{equation}
In this paper, we analyse the joint dynamics of two droplets and choose the dimensions of the periodic simulation box much larger than the droplets' radius and initial distance so that the periodic images of the droplets do not influence the results.

\section{Validation: axisymmetric collisions of two droplets}\label{validation}
The moving singularity model was validated for the axisymmetric collision of two active droplets, for which the exact dynamics was recently solved completely for various $\mbox{Pe}$~\cite{lippera2020collisions}.  In a head-on collision, droplets slow down and stop at a minimum distance $d_{\text{min}}$ from each other, as a result of their anti-chemotactic behaviour (they are effectively repelled by the solute they emit). The confinement-induced accumulation of the emitted solute between the droplets reverses the chemical polarity of the droplets which start swimming in the opposite direction and rebound.  

\begin{figure}[t]
\centering
\includegraphics[width=\columnwidth]{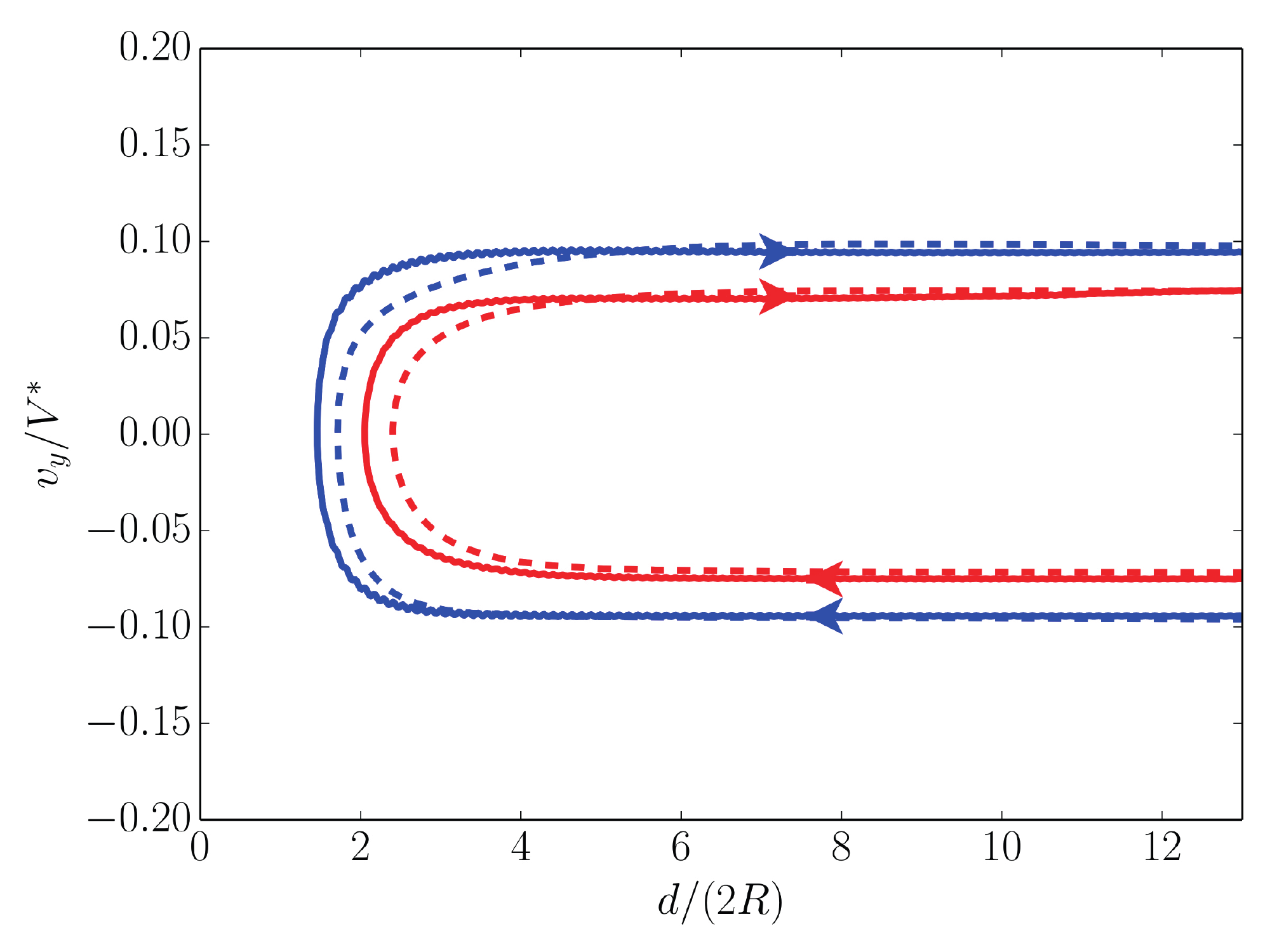}
\caption{Droplet axial velocity in an axisymmetric head-on collision with a second identical droplet for $\mbox{Pe=6}$ (red) and $\mbox{Pe}=8$ (blue) as obtained using the moving singularity model, Eqs.~\eqref{polarity}--\eqref{unst_diff} (solid) and from the exact result of the {fully-coupled} hydro-chemical model~\cite{lippera2020collisions} (dashed).}
\label{Headon}
\end{figure}

Figure~\ref{Headon} presents the evolution of the droplet velocity as a function of their relative distance as predicted by the moving singularity model and the exact solution~\citep{lippera2020collisions}. The droplets' slowing down and reversal dynamics is clearly visible and the moving singularity model is shown to provide a quantitatively accurate approximation of the exact rebound dynamics. The  rebound distance $d_{\text{min}}$ is slightly underestimated in the moving singularity model, which is consistent with the modelling of the droplet as point singularities for the chemical field which effectively reduces the confinement-induced accumulation of solute between the droplets. 

{\section{Three-dimensional collisions and alignment sensitivity}\label{app:3d}

For simplicity of analysis, and motivated by the two-dimensional kinematics of active droplets observed in experiments, the results presented in the main text are restricted to purely co-planar trajectories of the droplets. The sensitivity of the results presented to symmetric collisions in Section~\ref{SymColl} is analysed here by introducing a small three-dimensional perturbation $\delta$, defined as follows: the initial conditions are identical to that considered in Section~\ref{SymColl} (Figure~\ref{SchemaCollisions}a), but the droplets are now located on two different horizontal planes separated by a distance $\delta$ in the $z$-direction. 

\begin{figure}
\begin{center}
\includegraphics[width=0.5\textwidth]{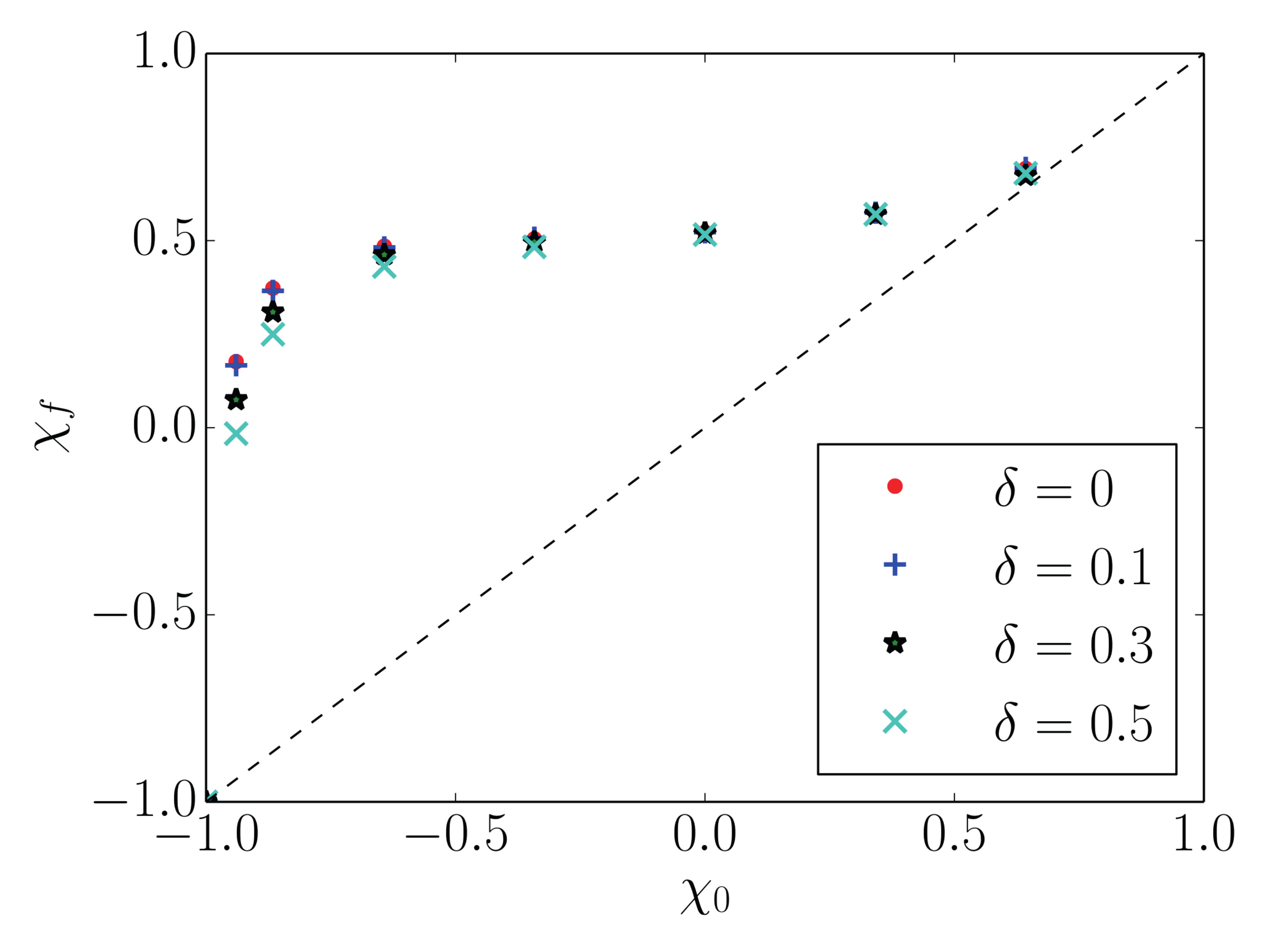}
\caption{{Sensitivity of the droplets' alignment in coplanar symmetric collisions (Figure~\ref{scattering_angle}) to a small initial misalignment $\delta$, defined as the initial vertical distance between the two droplets. The evolution of the final relative orientation $\chi_f$ with their initial relative heading $\chi_0$ is presented for $\Pe=6$ and $\delta=0$, $0.1$, $0.3$ and $0.5$. }}\label{scattering_angle_delta}
\end{center}
\end{figure}
We observe that a small perturbation of the co-planar configuration does not modify the main results and conclusions, neither qualitatively nor quantitatively, in particular the alignment property of the collision over a large range of incoming relative orientation. This is obviously not the case for purely head-on collisions, $\chi_0\approx -1$, for which a small vertical misalignment is essentially equivalent to a small variation of the initial relative cosine $\chi_0$, whose sensitivity has already been emphasized and discussed in the main text: as a result, even a small misalignment $\delta$ induces large modifications of the collision's outcome.}

\section{Simplified model for the evolution of the polarity evolution}
\label{reducedmodel}
For an isolated droplet in steady self-propulsion, its velocity $\bs v$ and polarity ${\bs\Pi=-\langle c\bs n\rangle}$ are proportional, see Eq.~\eqref{polarity}. During the collision with a second droplet, the self-propulsion still adjusts instantaneously to the chemical distribution in the Stokes regime, but the chemical polarity now results from the translation of the droplet and the unsteady diffusion of the chemical trail left behind it. This introduces a finite relaxation time $\tau$ of the self-polarity or wake toward ${(c^*R^2/2V^*)\bs v}$ (or to zero if the droplet stops moving), and a simple model for the polarity dynamics is an overdamped relaxation
\begin{align}
\frac{d \bs \Pi_i}{d t}=\frac{1}{\tau}\left(\frac{c^*R^2}{2V^*}\bs v_i-\bs \Pi_i \right).
\label{Pola_dynamics_approx}
\end{align}
As a result, the polarity magnitude ${\Pi_i=|\bs\Pi_i|}$ and direction ${\bs p_i=\bs\Pi_i/\Pi_i}$ satify
\begin{align}
\label{Polarity_dynamics_full1}
\frac{d \Pi_i}{d t}&=\frac{1}{\tau}\left(\frac{c^*R^2}{2V^*} \bs v_i\cdot\bs p_i- \Pi_i \right)\\
\label{Polarity_dynamics_full2}
\frac{d \bs p_i}{d t}&=\frac{c^*R^2}{2\tau V^*\Pi_i}(\bs p_i\times\bs v_i)\times \bs p_i.
\end{align}
In the following, we further neglect changes in magnitude of the polarity; as a consequence, the self-induced propulsion velocity (i.e. that due to the solute released by the droplet itself) has constant magnitude $v_\textrm{sp}$ and ${\Pi_i\approx (c^*R^2/2V^*)v_\textrm{sp}}$, and the wake's orientational dynamics simplifies into
\begin{align}
\frac{ d \bs p_i }{d t}&=\frac{(\bs p_i \times \bs v_i)\times \bs p_i}{\tau v_\textrm{sp}}
\end{align}
Note that neglecting changes in the polarity magnitude is only valid when the {chemical }wake reorganization is dominated by its reorientation (as in Figure~\ref{snapshots_collision}) and certainly does  not hold for purely head-on collisions where the polarity must vanish in magnitude in order to reverse direction~\cite{lippera2020collisions}.


\end{document}